\documentclass[12pt]{article}

\usepackage{amsmath, amssymb, graphicx, hyperref}
\usepackage{authblk}
\usepackage{geometry}
\usepackage{siunitx}
\usepackage{wrapfig}
\usepackage{caption}
\usepackage{xcolor}
\newcommand{\hide}[1]{}

\title{\textbf{A Multimodal Vision Transformer-based Modeling Framework for Prediction of Fluid Flows in Energy Systems}}

\author[1]{Kiran Yalamanchi}
\author[1]{Shivam Barwey}
\author[1]{Ibrahim Jarrah}
\author[1]{Pinaki Pal}

\affil[1]{\textit{Transportation and Power Systems Division, Argonne National Laboratory, Lemont, IL, USA}}

\date{}

\begin{document}

\maketitle

\begin{abstract}
Computational fluid dynamics (CFD) simulations of complex fluid flows in energy systems are prohibitively expensive due to strong nonlinearities and multiscale-multiphysics interactions. In this work, we present a transformer-based modeling framework for prediction of fluid flows, and demonstrate it for high-pressure gas injection phenomena relevant to reciprocating engines. The approach employs a hierarchical Vision Transformer (SwinV2-UNet) architecture that processes multimodal flow datasets from multi-fidelity simulations. The model architecture is conditioned on auxiliary tokens explicitly encoding the data modality and time increment. Model performance is assessed on two different tasks: (1) spatiotemporal rollouts, where the model autoregressively predicts the flow state at future times; and (2) feature transformation, where the model infers unobserved fields/views from observed fields/views. We train separate models on multimodal datasets generated from in-house CFD simulations of argon jet injection into a nitrogen environment, encompassing multiple grid resolutions, turbulence models, and equations of state. The resulting data-driven models learn to generalize across resolutions and modalities, accurately forecasting the flow evolution and reconstructing missing flow-field information from limited views. This work demonstrates how large vision transformer-based models can be adapted to advance predictive modeling of complex fluid flow systems.
\end{abstract}

\section{Introduction}
Capturing the spatiotemporal evolution of complex fluid flows in energy and propulsion systems is challenging. For instance, accurately predicting the dynamics of high-pressure gas jets, liquid sprays, and turbulent reacting flows is critical for designing efficient fuel injectors and combustion engines. Predicting such fluid dynamics phenomena using traditional computational fluid dynamics (CFD) simulations is often prohibitively expensive due to the inherent nonlinear, multiphysics, and multiscale behavior. This motivates the development of data-driven surrogate modeling approaches. In particular, the emerging paradigm of scientific foundation models – large-scale pretrained neural networks – offers a promising avenue: by leveraging vast heterogeneous datasets, these models can learn general representations that can be fine-tuned or adapted to multiple tasks and regimes.

Traditional scientific machine learning methods for PDEs and fluid flows provide important context. Neural operators~\cite{kovachki2023neural}~\cite{li2021neural}, including DeepONet~\cite{lu2021learning} and Fourier Neural Operators~\cite{li2020fourier}, have shown that neural networks can learn to map initial conditions or parameters to flow solutions efficiently by learning mappings between function spaces, enabling efficient solution of parametric PDEs through operator learning. However, these models are typically trained per-application: one network per geometry or flow configuration. This specialization limits their ability to generalize to new conditions or physics. To overcome this, the notion of foundation models for PDEs has emerged. Instead of isolating individual problems, a foundation model is trained across many simulations, with different equations or parameters, to learn a unified operator. In this view, foundation models for PDEs aim to approximate solution operators for large classes of PDEs within a single DNN and learn a general operator that can represent and infer the forward dynamics of distinct physical systems~\cite{zhang2024prose,subramanian2023towards}. By exposing the network to a variety of flow regimes, a single foundation model can, in principle, capture shared structures and generalize beyond the training set. This approach is analogous to how large language models acquire broad knowledge by pretraining on diverse text corpora.

Foundation modeling ideas have already begun to transform the domain areas of earth systems and fluid science. In climate and weather modeling, large pretrained transformers have delivered impressive gains. The ClimaX model~\cite{nguyen2023climax} was pretrained on heterogeneous climate simulations (CMIP6 data) and then fine-tuned to forecasting and projection tasks; it achieved superior accuracy even when deployed on resolutions and variables it did not see during training. Scaling transformer neural networks~\cite{nguyen2024scaling} has demonstrated skillful and reliable medium-range weather forecasting capabilities. FourCastNet~\cite{pathak2022fourcastnet} demonstrated the effectiveness of adaptive Fourier neural operators for global weather forecasting, achieving high-resolution predictions with improved computational efficiency. Beyond the atmosphere, researchers have explored multimodal sensor fusion: for instance, remote sensing foundation models~\cite{conti2025advancing} ingest optical, infrared, and radar imagery from disparate satellites to build versatile feature extractors for earth observations. These works demonstrate that vision-transformer architectures can assimilate spatio-temporal data across variables and scales.

In parallel, specialized foundation models have emerged for fluid PDEs. Poseidon~\cite{herde2024poseidon} employed a multiscale operator transformer pretrained on a diverse dataset of 2D flow simulations, enabling it to efficiently solve new PDEs and outperform operator-learning baselines on numerous benchmarks. Similarly, PROSE-FD~\cite{zhang2024prose}, a multimodal foundation model, was trained on a mixture of incompressible and compressible Navier–Stokes and shallow-water systems; by fusing symbolic physical information, it achieved strong zero-shot generalization to new flow scenarios. VICON~\cite{cao2024vicon} introduced vision in-context operator networks that efficiently process 2D functions through patch-wise operations, demonstrating superior performance and computational efficiency for compressible flow predictions. Multiple Physics Pretraining (MPP)~\cite{mccabe2023mpp} employed axial attention mechanisms for physical surrogate models, further demonstrating the versatility of transformer architectures for scientific computing. PhysiX~\cite{nguyen2025physix} introduced a large-scale autoregressive generative model for physics simulations, using discrete tokenization to encode physical processes at different scales and demonstrating effective knowledge transfer from natural videos to physics simulations. EddyFormer~\cite{du2025eddyformer} presented a Transformer-based spectral-element architecture for large-scale turbulence simulation, achieving DNS-level accuracy at 256$^3$ resolution with 30$\times$ speedup while preserving physics-invariant metrics across domains. These studies demonstrate that transformer architectures can learn complex fluid solution operators across varied equations, geometries, and boundary conditions.

While recent foundation models for fluid dynamics have demonstrated strong operator-learning capabilities, their focus has largely been on learning idealized PDE mappings, typically single-phase and well-posed systems. These studies emphasize the mathematical generality of PDE solvers rather than the data and modality heterogeneity that characterizes applied fluid problems. In this context, the present work explores a complementary challenge: developing a unified framework that can integrate and reason across heterogeneous observational modalities and simulation-derived flow fields within realistic engineering regimes. Specifically, we target compressible jet flows relevant to propulsion and energy systems. We develop a multimodal vision transformer-based modeling framework built on a SwinV2 Transformer–based encoder–decoder U-Net~\cite{liu2022swin}. We further embed contextual information— dataset resolution, modality, turbulence closure, and temporal increment—enabling transformer-based models sharing this architecture to adapt their behavior across diverse data sources. Drawing inspiration from hierarchical Swin architectures like Aurora~\cite{bodnar2025aurora}, our framework is designed to perform multiple tasks: (i) autoregressive spatiotemporal rollouts for forecasting flow evolution, and (ii) feature transformations such as reconstructing unobserved velocity fields from density fields.

In this work, we train our framework on a curated set of multi-fidelity CFD simulations of an argon jet injected into a quiescent nitrogen environment, a non-combusting analog for gaseous fuel injection in piston engines. They span coarse and fine grids, RANS and LES turbulence closures, and ideal vs. real gas equations of state. The diverse simulation data enables training separate transformer-based models sharing a common architecture on multiple physics representations in a consistent way. The trained transformer-based models are evaluated for future-state prediction and cross-modal field reconstruction tasks. Our results show that the transformer-based models successfully generalize across grid resolutions, turbulence models, and data modalities. In summary, we present a proof-of-concept multimodal transformer-based modeling framework that bridges modern transformer ideas with fluid-dynamics data relevant to practical energy systems, paving the way for fast, data-driven surrogate models of complex flows in engineering applications.

\section{Data Curation}
All training and evaluation conducted in this work are based on a single, well-characterized physical configuration: an argon jet issued at \SI{35}{bar} into a quiescent nitrogen environment at \SI{5}{bar}. This configuration provides a controlled test bed for assessing data-driven flow modeling under thermodynamically complex conditions. Numerical realizations of the flow were generated using the \textsc{CONVERGE} CFD solver~\cite{senecal2007converge}. Two grid resolutions were employed: a \emph{coarse} grid and a \emph{fine} grid. Across these grids, multiple modeling strategies were applied to introduce systematic physical and numerical variability. Both the turbulence closure model —Reynolds-Averaged Navier–Stokes (RANS) and Large-eddy simulation (LES)—and thermodynamic treatment—ideal-gas (IG) and real-gas (RG) equations of state—were varied. Within the RANS framework, additional simulations were performed with modified effective Schmidt numbers ($Sc$) to examine the influence of differential diffusion on scalar transport. These variations yield a dataset spanning different levels of physical fidelity and numerical resolution, enabling robust multi-fidelity learning.

\begin{table}[htbp]
    \centering
    \caption{Curated simulation datasets for argon (\SI{35}{bar}) injection into nitrogen (\SI{5}{bar}). Each case label encodes grid resolution (C/F), turbulence model, equation of state, and Schmidt number (if applicable).}
    \label{tab:curated_datasets}
    \vspace{4pt}
    \begin{tabular}{lcccc}
        \hline
        \textbf{ID} & \textbf{Grid} & \textbf{Turbulence} & \textbf{Equation of State} & \textbf{$Sc$} \\
        \hline
        C--LES--RG         & Coarse & LES  & Real gas  & --   \\
        C--RANS--RG        & Coarse & RANS & Real gas  & --   \\
        C--RANS--RG--0.5   & Coarse & RANS & Real gas  & 0.50 \\
        F--LES--IG         & Fine   & LES  & Ideal gas & --   \\
        F--RANS--IG        & Fine   & RANS & Ideal gas & --   \\
        F--RANS--RG        & Fine   & RANS & Real gas  & --   \\
        F--RANS--RG--0.25  & Fine   & RANS & Real gas  & 0.25 \\
        \hline
    \end{tabular}
\end{table}


Each simulation shares identical inlet and boundary conditions to isolate the effects of the modeling choices outlined above. The resulting dataset comprises seven distinct flow cases, summarized in Table~\ref{tab:curated_datasets}. Each case is encoded using a concise label reflecting grid resolution (C/F), turbulence model (RANS/LES), equation of state (IG/RG), and, where applicable, the specified Schmidt number. For every case, six scalar or vector fields are exported at uniform physical time intervals: Argon partial density~($\rho$), argon mass fraction~($Y_{\text{Ar}}$), turbulent kinetic energy~($k$), and the three velocity components~($u,v,w$). For the purpose of testing multimodal learning capabilities, three sets of 2D fields are derived from the 3D data. The following terminology is adopted to refer to these throughout the subsequent sections. \emph{Longitudinal slice} refers to flow-field snapshots on the horizontal ($x-z$) cut-plane through the jet centerline ($y=0$), where $z$ is the streamwise direction. \emph{Longitudinal projection} refers to line-of-sight--integrated projected flow-field snapshots (analogous to X-ray radiography measurements) on the same plane, obtained by integrating along~$\pm y$ through the domain. \emph{Transverse slice} corresponds to density field snapshots (analogous to X-ray tomography measurements) corresponding to two axial planes located at $z=\SI{2}{mm}$ and $z=\SI{10}{mm}$ from the injector. Together, these views give distinct orientations relative to the jet.

\section{Methodology}
Our framework uses a single Transformer-based backbone to address two related tasks. The first task is spatiotemporal prediction: given the flow state at time $t$, the model autoregressively advances to $t+\Delta t$ by predicting the residual $\Delta\mathbf{u}$, which is then used to compute $\mathbf{u}_{t+\Delta t}=\mathbf{u}_t+\Delta\mathbf{u}$. The second task is feature transformation within a single time slice: given a subset of variables or measurements at time $t$, the model infers other target fields at that same time. For example, the network learning to map a longitudinal projected density field to a corresponding longitudinal or transverse slice density field. Importantly, these two tasks share the same core architecture; the only differences lie in the input conditioning. In the following we first describe our Transformer backbone (used for temporal rollouts) and then explain how it is adapted for feature-transformation tasks.

\subsection{Model Architecture}
The model backbone uses a hierarchical vision transformer inspired by recent SwinV2 designs~\cite{liu2022swin}. The architecture begins with patch embedding: each input field (which may have multiple channels) is partitioned into a grid of non-overlapping patches using a convolutional layer with kernel size and stride equal to the patch size. Specifically, for an input tensor of shape $[B, C_{\text{in}}, H, W]$ where $B$ is the batch size, $C_{\text{in}}$ is the number of input channels, and $H \times W$ is the spatial resolution, a $2D$ convolution with kernel size $(P_h, P_w)$ and stride $(P_h, P_w)$ produces patches of size $P_h \times P_w$. The convolution maps $C_{\text{in}}$ input channels to an embedding dimension $C$ (base embedding dimension), resulting in a tensor of shape $[B, C, H/P_h, W/P_w]$. This is then permuted to $[B, H/P_h, W/P_w, C]$ and normalized using LayerNorm along the channel dimension, producing patch embeddings (also referred to as token embeddings) of shape $[B, H', W', C]$ where $H' = H/P_h$ and $W' = W/P_w$ are the token grid dimensions. Positional information is implicitly encoded through the spatial arrangement of patches, as the shifted window attention mechanism in SwinV2 inherently captures relative spatial relationships. In a vanilla ViT, global self-attention is applied across all patches, allowing long-range spatial dependencies to be captured; however, ViT produces a single coarser-resolution feature map and its global attention has quadratic cost in the number of patches. This can become impractical for the high-resolution fluid datasets. Physics-informed neural operators~\cite{li2024physics,wang2021learning} have shown promise for incorporating physical constraints, though our approach focuses on learning from diverse simulation data rather than explicit physics constraints.

To address these issues, we employ a Shifted Window Transformer (SwinV2) architecture~\cite{liu2022swin}. After the initial patch embedding, the model applies a series of SwinV2 Transformer blocks organized in a hierarchical encoder-decoder structure. In each SwinV2 block, self-attention is restricted to small, non-overlapping spatial windows of patches, which greatly reduces computation from quadratic to linear in the number of patches. Crucially, the window configuration is shifted between successive blocks (e.g., half-window offset), enabling information to flow across window boundaries while still keeping each attention operation local. This SwinV2 backbone is implemented in a U-shaped encoder–decoder layout (SwinV2-UNet). The encoder builds a multi-resolution hierarchy through patch merging. After SwinV2 blocks at one resolution level, a patch-merging layer operates on groups of $2 \times 2$ adjacent patch tokens. Each patch token has dimension $C_i$ at stage $i$. The four adjacent patches are concatenated along the channel dimension, resulting in a $4C_i$-dimensional vector. This concatenated vector is then passed through a linear projection layer that reduces the dimension to $2C_i$, effectively halving the spatial resolution from $H_i \times W_i$ to $H_i/2 \times W_i/2$ while doubling the channel dimension from $C_i$ to $2C_i$. The encoder consists of two downsampling stages: each stage applies SwinV2 Transformer blocks, followed by a residual connection, then a ConvNeXt block~\cite{liu2022convnext} (for skip connections), and finally patch-merging (except at the last stage). The ConvNeXt blocks apply depthwise separable convolutions and layer normalization to extract local spatial features, providing a convolutional inductive bias that complements the transformer's global attention mechanism. The outputs of these ConvNeXt blocks are stored for skip connections to the decoder. This merging yields features at progressively coarser scales with increasing channel capacity, enabling the model to capture multi-scale context while maintaining linear complexity in the input size.

\begin{figure}[htbp]
    \centering
    \includegraphics[width=1\textwidth]{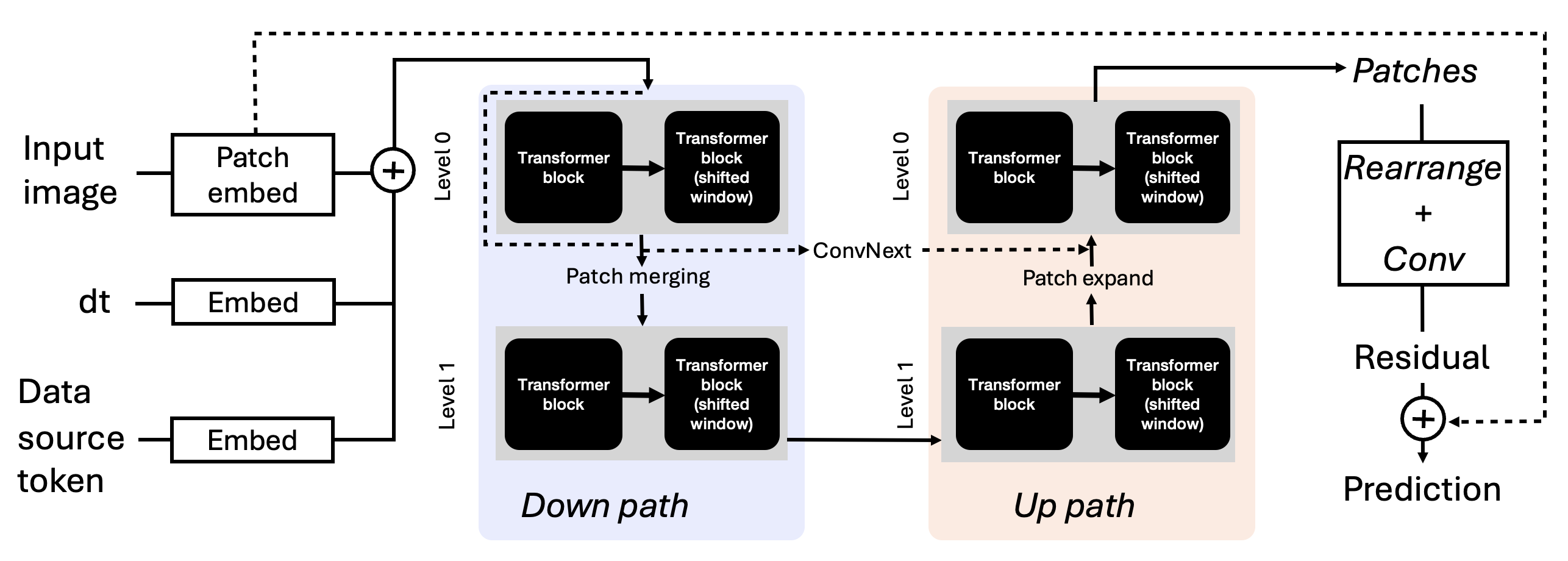}
    \caption{Schematic of the SwinV2-UNet architecture used for spatiotemporal prediction.}
    \label{fig:1}
\end{figure}

The decoder mirrors the encoder with two upsampling stages. In each decoding stage, SwinV2 blocks process the features, followed by patch expansion layers (the inverse of patch merging) that expand each token into a higher-resolution group of tokens. Patch expansion operates by first projecting each token to $4C_{\text{out}}$ dimensions where $C_{\text{out}} = C_{\text{in}}/2$, then rearranging to double the spatial dimensions while halving the channels from $C_{\text{in}}$ to $C_{\text{out}}$. Skip connections link encoder and decoder stages at matching resolutions: the ConvNeXt outputs from the encoder are added to the corresponding decoder features after patch expansion, preserving fine spatial detail. After the final decoder stage, an unpatch-and-convolution layer converts the token embeddings back to image space. For input tokens of shape $[B, H', W', C]$ where $H' = H/P_h$ and $W' = W/P_w$ are the token grid dimensions and $C$ is the embedding dimension, each token is linearly projected to $C \cdot P_h \cdot P_w$ dimensions. The rearranging operation then reshapes from $[B, H', W', C \cdot P_h \cdot P_w]$ to $[B, C, H, W]$, reconstructing the full spatial resolution by expanding each token into its $P_h \times P_w$ patch. A final convolution layer maps from the embedding dimension $C$ to the output channels $C_{\text{out}}$, producing the output field of shape $[B, C_{\text{out}}, H, W]$ matching the input spatial resolution. This design combines the long-range, multi-scale context of the SwinV2 encoder with the high-resolution reconstruction capability of a U-shaped network. A schematic of this SwinV2-UNet architecture is shown in Figure~\ref{fig:1} for spatiotemporal prediction, which consists of ~ 5 million parameters.

For the spatiotemporal prediction task, the loss function is defined as the mean squared error (MSE) between the predicted flow fields and the ground-truth fields. Specifically, for a prediction $\hat{\mathbf{u}}_{t+\Delta t}$ and ground truth $\mathbf{u}_{t+\Delta t}$, the loss is:
\begin{equation}
\mathcal{L}_{\text{temporal}} = \frac{1}{N} \sum_{i=1}^{N} \|\hat{\mathbf{u}}_{t+\Delta t}^{(i)} - \mathbf{u}_{t+\Delta t}^{(i)}\|_2^2,
\end{equation}
where $N$ is the number of spatial locations and $\|\cdot\|_2$ denotes the $L_2$ norm. In this study, multiple training strategies are explored. For multi-step rollout training, the loss is accumulated across all time steps in the rollout horizon and averaged. When using the pushforward training strategy~\cite{lee2023autoregressive,list2025differentiability}, the loss is computed only on the final time step of the rollout to reduce computational cost from incorporating intermediate steps. On the other hand, for feature transformation tasks, we use either MSE loss or Euclidean loss depending on the specific case, as described in Section 4.2.

\subsection{Auxiliary Embeddings}
In addition to the patch tokens, we provide the model with auxiliary embeddings to condition its predictions on temporal and contextual factors. For the spatiotemporal prediction task, after patch embedding produces tokens of shape $[B, H', W', C]$, two auxiliary embeddings are computed and added element-wise to all patch tokens. A time-step embedding encodes the physical interval $\Delta t$ between consecutive flow states. The scalar $\Delta t$ (shape $[B, 1]$) is passed through a linear layer followed by LayerNorm to produce a vector embedding of dimension $C$ (shape $[B, C]$), which is then unsqueezed to $[B, 1, 1, C]$ and broadcast across all spatial locations, enabling the model to account for varying temporal scales. To incorporate dataset-specific context, a data source token (DST) embedding is computed. This embedding is derived from an eight-dimensional one-hot vector (shape $[B, 8]$) representing key attributes of the data sample: spatial resolution (fine or coarse), data modality (longitudinal slice, longitudinal projected or transverse slice at 2~mm or 10~mm), and simulation fidelity (LES or RANS). The one-hot vector is linearly projected to the embedding dimension $C$ followed by LayerNorm (producing shape $[B, C]$), then unsqueezed and broadcast to $[B, 1, 1, C]$ and added to all patch tokens, allowing the model to differentiate between datasets and adapt its feature processing accordingly.

For feature-transformation tasks, where both input and output correspond to the same time instance, the temporal embedding is omitted while the data source token remains. In this case, the token encoding is adjusted to reflect the specific input variable (e.g., no modality token if all the inputs are longitudinal projections). This flexible embedding strategy enables the shared Transformer backbone to generalize across multiple data configurations by modifying only the auxiliary token encodings. In addition, unlike the autoregressive rollout mode, the feature-transformation branch predicts the target field directly rather than as a residual correction to the input.

\section{Model Evaluation Studies}
The multimodal transformer-based modeling framework is evaluated on two primary categories of scientific learning tasks: (i) spatiotemporal prediction, in which the model autoregressively predicts the temporal evolution of multi-physics flow fields, and (ii) feature transformation, in which the model infers one set of physical variables or modalities from another at the same time instant. Both tasks leverage the same SwinV2-UNet Transformer backbone described in Section 3.1, trained with task-specific conditioning and loss formulations. The common objective is to assess the model's capability to learn physically consistent representations across time, space, and modality, using the datasets curated from multi-fidelity simulations.

The first category, spatiotemporal prediction, evaluates the model's capacity to propagate flow fields forward in time given an initial state and a prescribed temporal increment $\Delta t$. This task is critical for developing surrogate models that can emulate computationally expensive simulations at orders-of-magnitude faster runtimes. The second category, feature transformation, probes the model's ability to infer latent or missing information within a given time slice—such as reconstructing unmeasured velocity fields from density data or mapping between different modalities. Together, these tasks serve as complementary assessments of both the temporal and representational generalization capabilities of the model.

\subsection{Spatiotemporal Prediction}

For spatiotemporal prediction, the model is trained to capture the temporal evolution of flow-field. Given a snapshot of the flow state at time $t$, the network predicts the flow-field at next time step $t + \Delta t$ using an autoregressive formulation. Three rollout configurations are considered in this work. The first, termed \emph{single-step training}, involves predicting one future frame at a time, where ground-truth data at each step are used as inputs for the next prediction. The second, \emph{multi-step rollout training}, extends this approach by unrolling the model predictions over multiple consecutive time steps within a single training iteration, allowing gradients to propagate across the entire rollout horizon. This setup enables the network to learn stable long-term dynamics and reduces short-horizon overfitting. Finally, \emph{pushforward rollout training} still autoregressively advances the state over a rollout horizon during training, but intermediate steps are taken without gradient flow through them; only the terminal state of the rollout horizon is compared to ground truth for computing loss. So, the model is trained on long-horizon error, while avoiding backpropagation through the full recurrent chain. The results of these configurations are analyzed to evaluate predictive stability, temporal coherence, and physical fidelity.

\begin{figure}[htbp]
    \centering
    \includegraphics[width=1\textwidth]{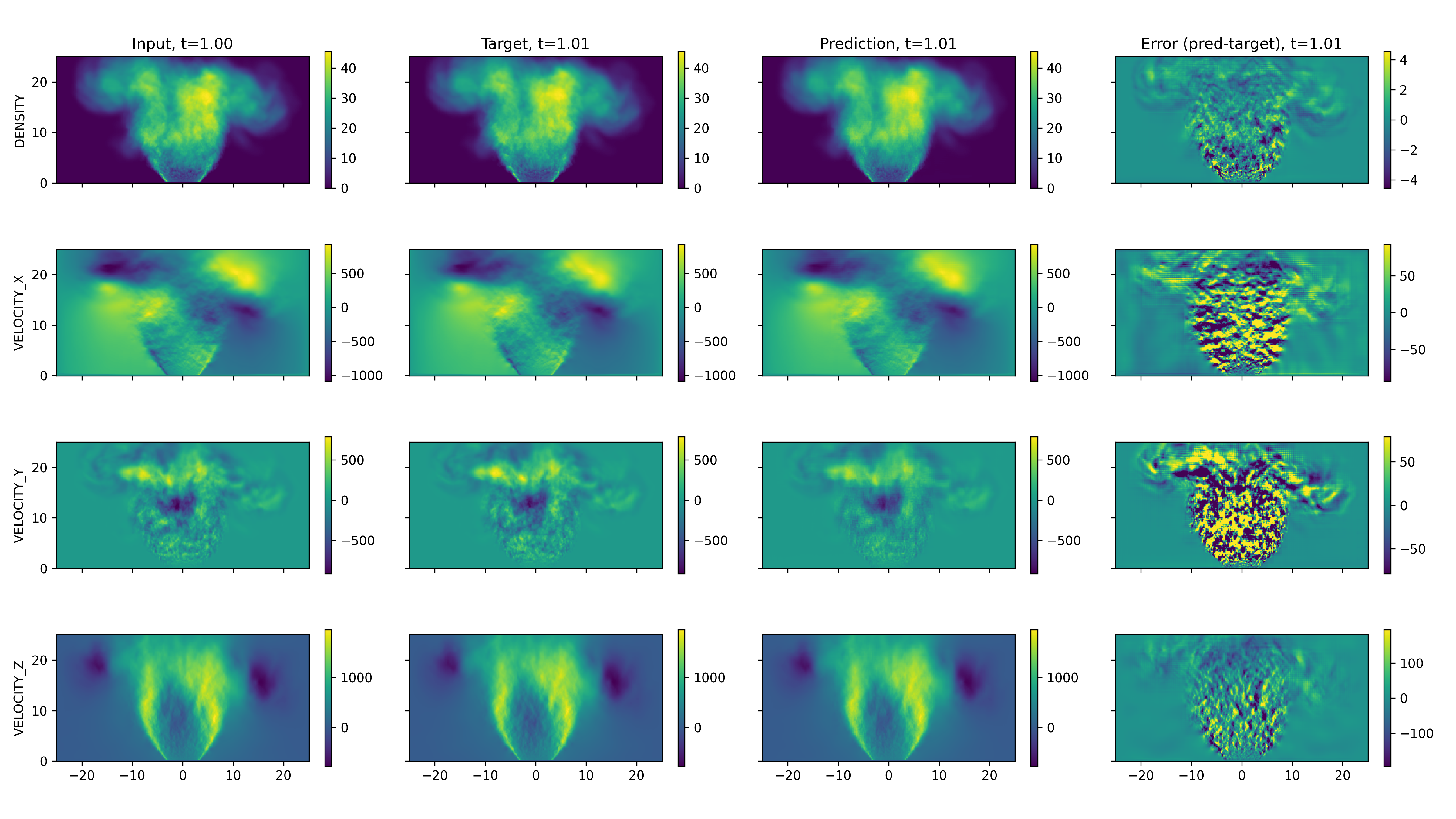}
    \caption{Spatiotemporal prediction results for longitudinal projected flow variables. Each row corresponds to a physical variable ($\rho$, $u$, $v$, $w$). Columns show: input at $t=1.00$~s, ground-truth target at $t=1.01$~s, baseline model (trained with single-step rollout) prediction at $t=1.01$~s, and local prediction error at $t=1.01$~s.}
    \label{fig:2}
\end{figure}

\begin{figure}[htbp]
    \centering
    \includegraphics[width=1\textwidth]{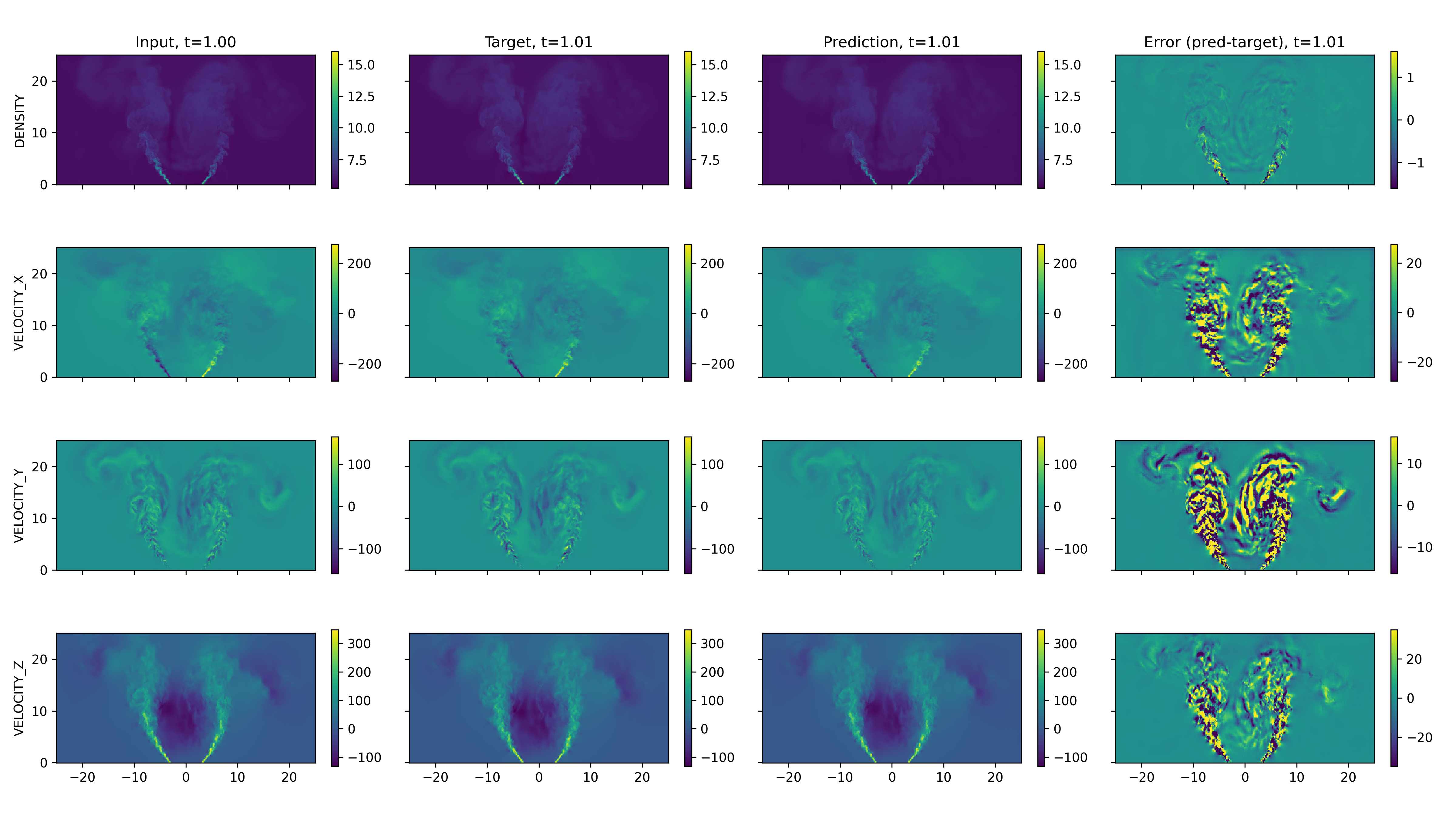}
    \caption{Spatiotemporal prediction results for longitudinal slice variables. Layout as in Figure~\ref{fig:2}, with rows for density and velocity components ($u$, $v$, $w$) on planar slices.}
    \label{fig:3}
\end{figure}

\begin{figure}[htbp]
    \centering
    \includegraphics[width=1\textwidth]{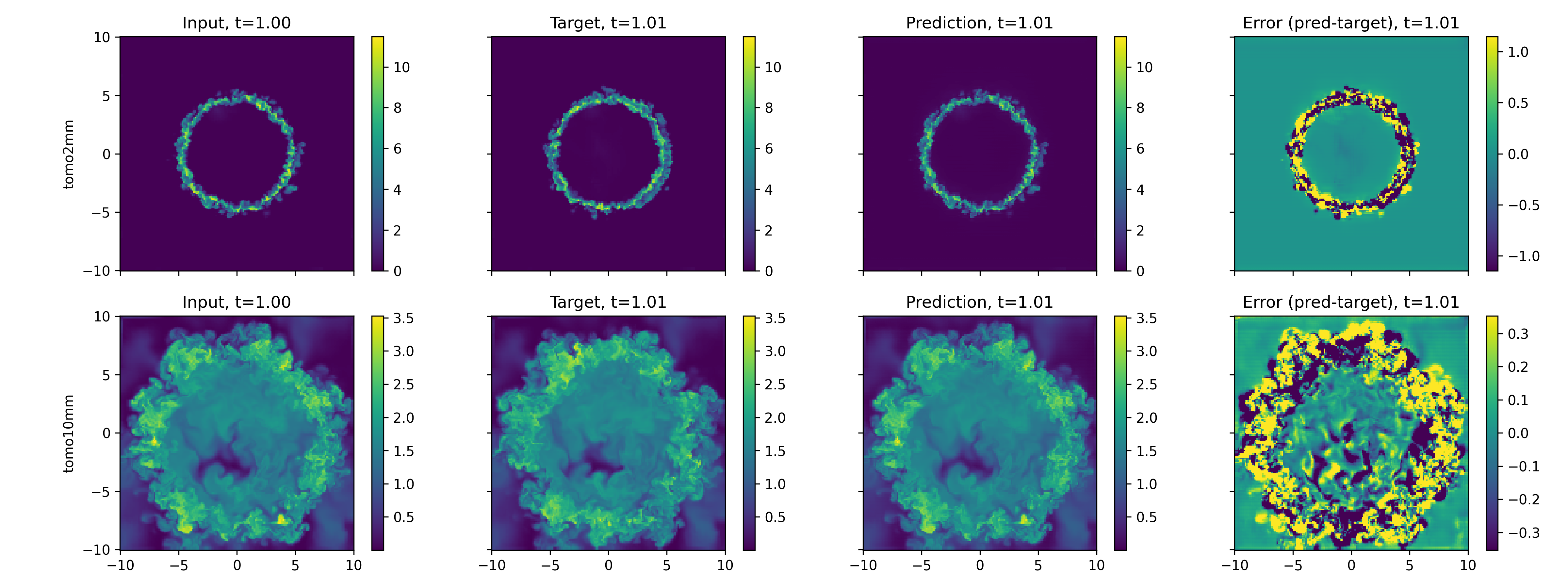}
    \caption{Spatiotemporal prediction results for density on transverse planes at $z=2$~mm and $z=10$~mm. Columns show input, target, baseline model prediction, and local prediction error.}
    \label{fig:4}
\end{figure}

The results in Figures~\ref{fig:2}--\ref{fig:4} illustrate the baseline model's performance on the spatiotemporal prediction task for longitudinal projected variables, longitudinal slice variables, and transverse slice densities, respectively. The baseline model corresponds to the single-step training configuration described above. The trained model's predictions are shown for the fine-grid LES ideal-gas case (F--LES--IG, Table~\ref{tab:curated_datasets}), which was not part of training dataset, for a single step from $t=1.00$~s to $t=1.01$~s. Across all modalities, the baseline model's predicted flow-fields accurately capture the large-scale evolution and edge transitions of the spray, as evidenced by low error regions along the moving front. However, the error maps reveal that finer-scale, intrinsic details within the flow are not fully captured, with significant discrepancies concentrated in regions of complex internal structure. This pattern is consistent for both longitudinal projected and slice variables, as well as for transverse slice density. The baseline model's ability to preserve sharp interfaces and global motion is robust, but the recovery of small-scale features remains challenging.

\begin{figure}[p]
    \centering
    \rotatebox{90}{%
        \begin{minipage}{1\textheight}
            \centering
            \includegraphics[width=1\textheight]{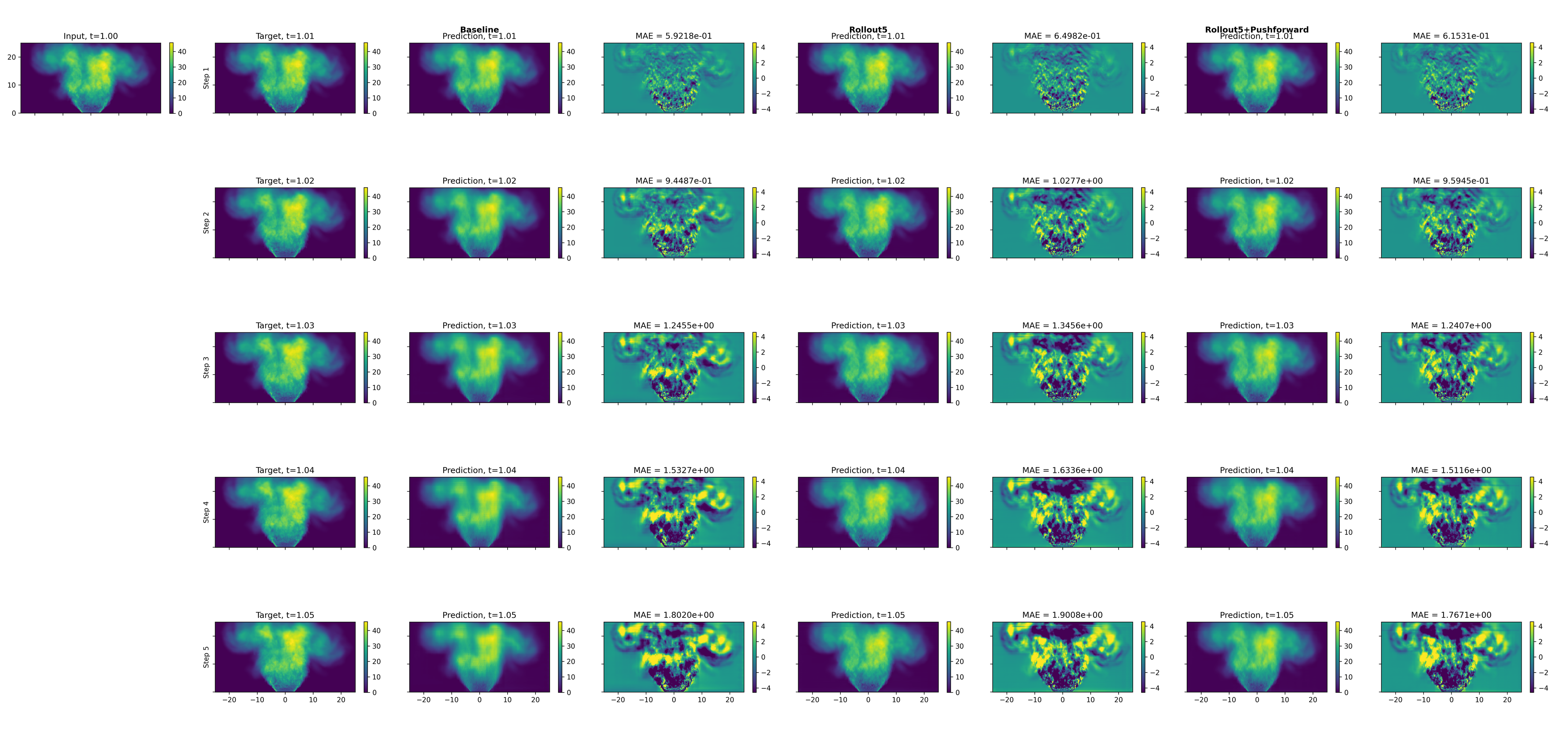}
            \vspace{0.5em}
            \caption{Multi-step rollout prediction for longitudinal projected density. Each row corresponds to an autoregressively predicted time step (from 1 to 5 steps ahead). Columns show: input, target, prediction/error for baseline single-step model), prediction/error for multi-step rollout model, prediction/error for multi-step rollout with pushforward. Mean absolute error (MAE) is indicated in error subplots.}
            \label{fig:5}
        \end{minipage}%
    }
\end{figure}

\begin{figure}[p]
    \centering
    \rotatebox{90}{%
        \begin{minipage}{1\textheight}
            \centering
            \includegraphics[width=1\textheight]{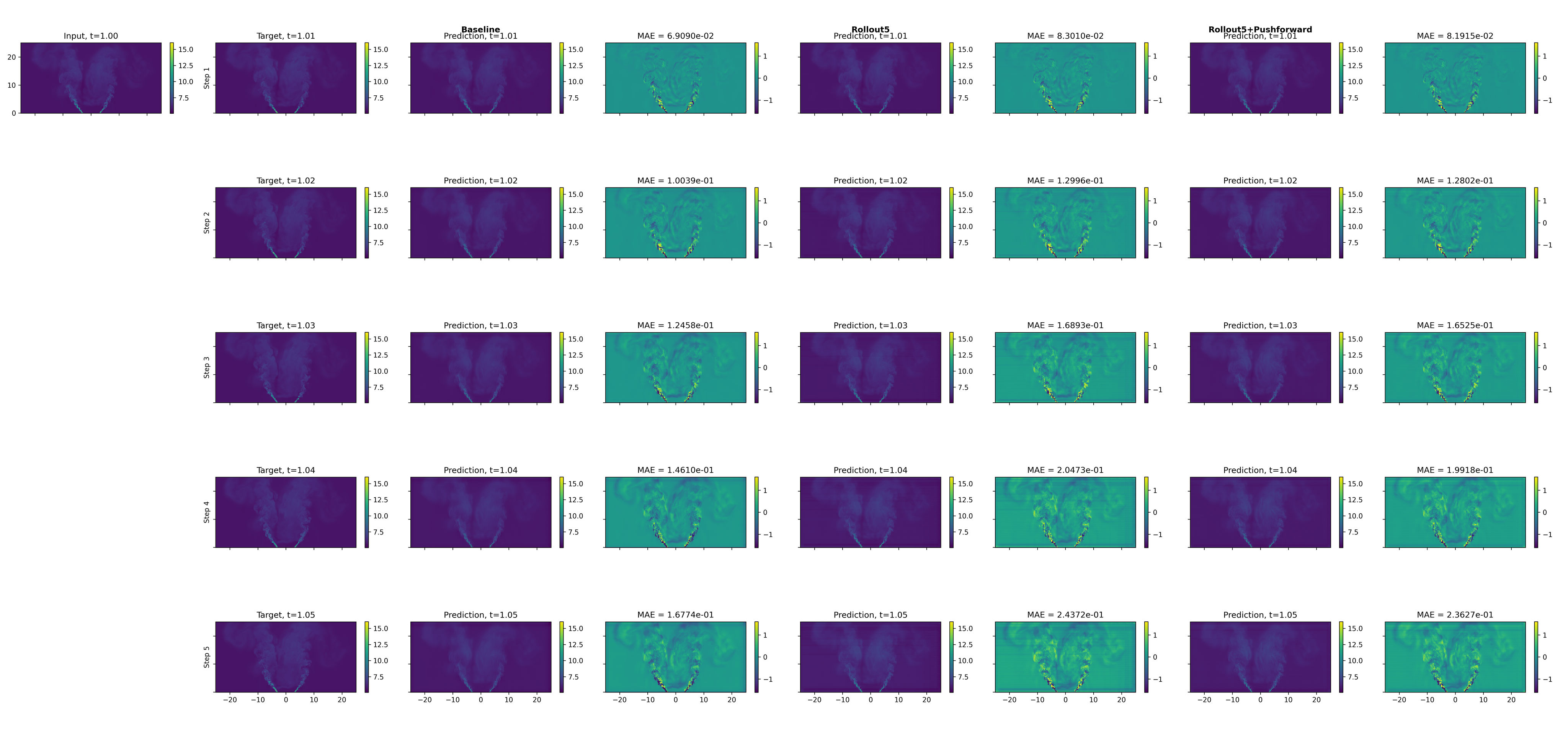}
            \vspace{0.5em}
            \caption{Multi-step rollout prediction for longitudinal slice density. Layout as in Figure~\ref{fig:5}, with rows for each autoregressive time step.}
            \label{fig:6}
        \end{minipage}%
    }
\end{figure}

\begin{figure}[p]
    \centering
    \rotatebox{90}{%
        \begin{minipage}{1\textheight}
            \centering
            \includegraphics[width=1\textheight]{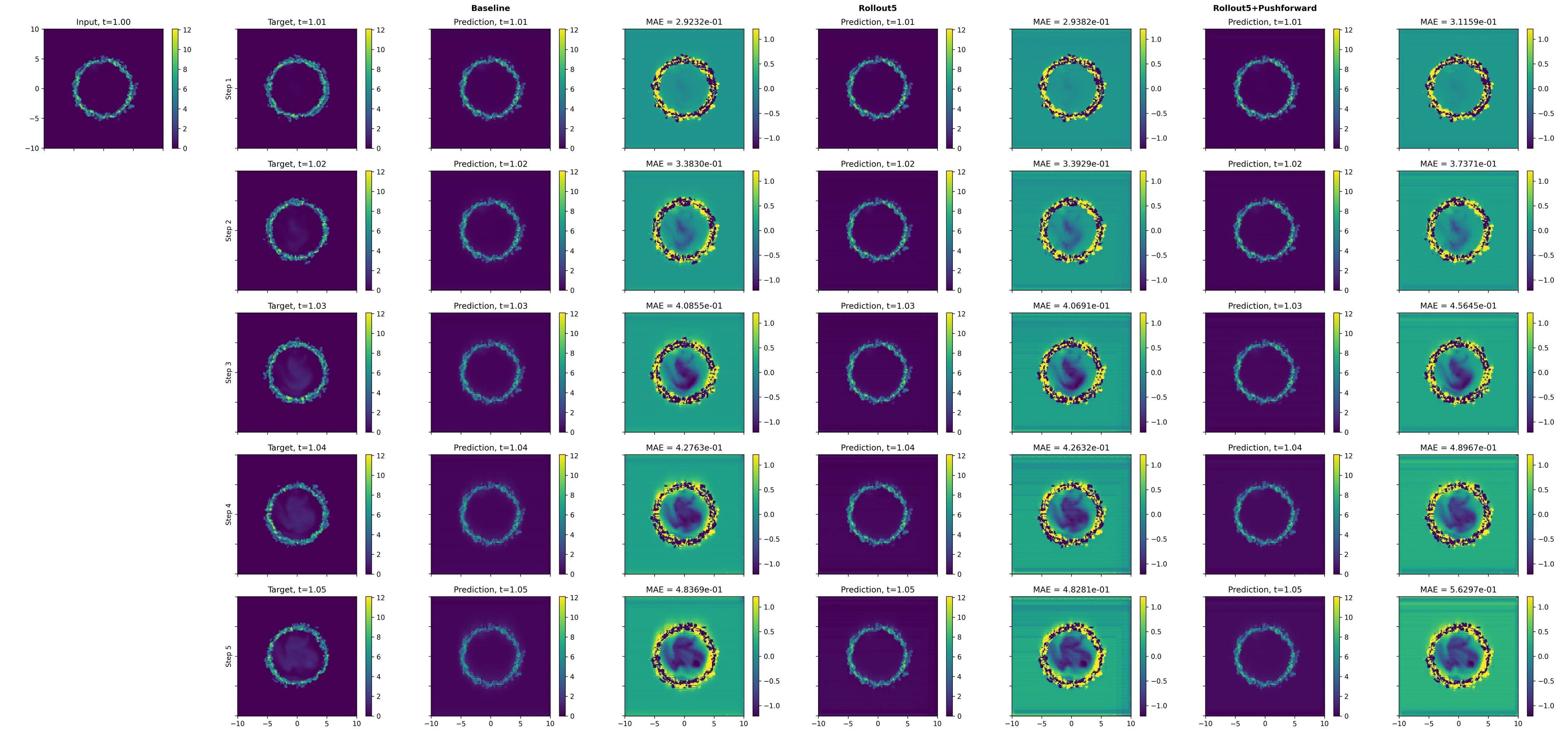}
            \vspace{0.5em}
            \caption{Multi-step rollout prediction for transverse slice density at $z=2$~mm. Layout same as in Figure~\ref{fig:5}.}
            \label{fig:7}
        \end{minipage}%
    }
\end{figure}

\begin{figure}[p]
    \centering
    \rotatebox{90}{%
        \begin{minipage}{1\textheight}
            \centering
            \includegraphics[width=1\textheight]{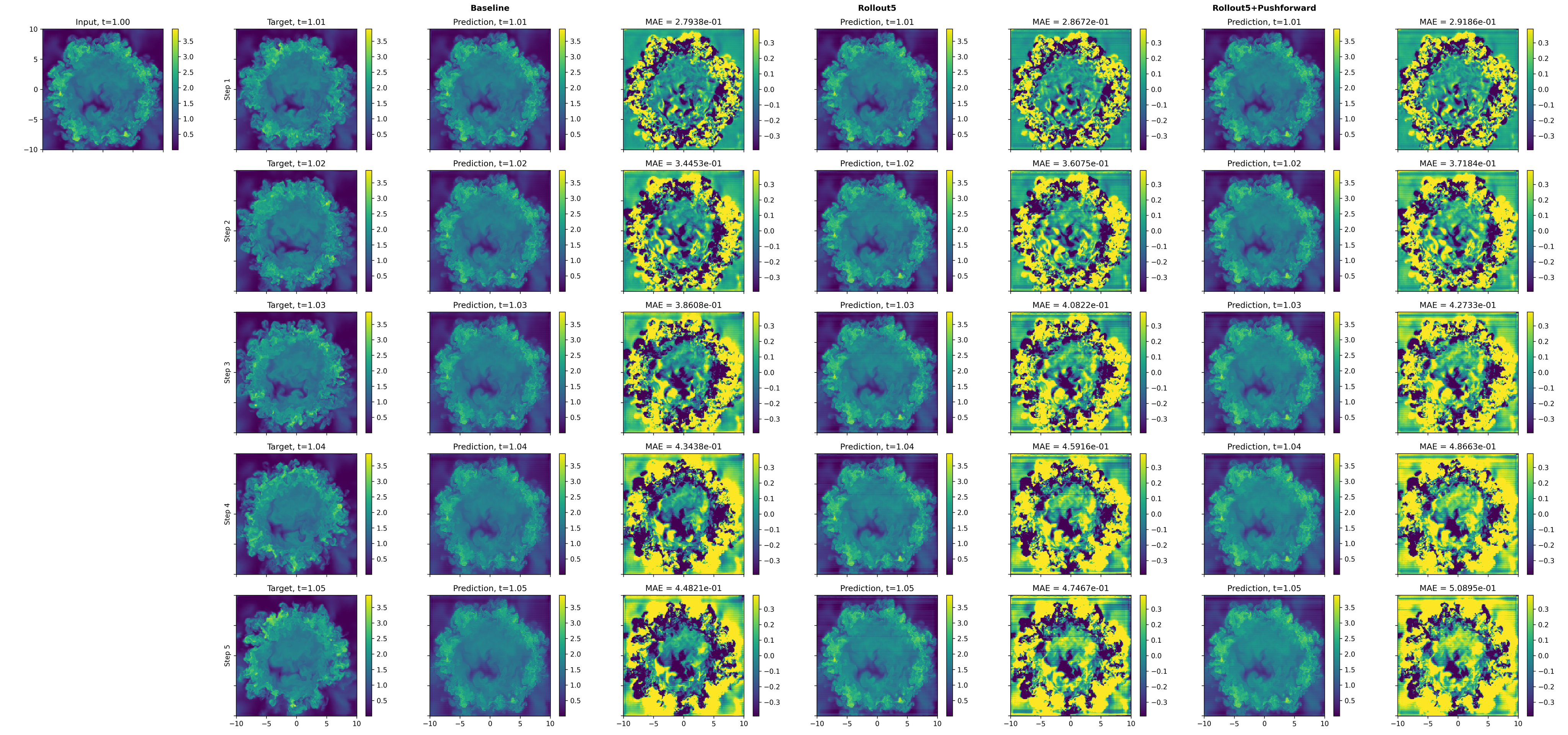}
            \vspace{0.5em}
            \caption{Multi-step rollout prediction for transverse slice density at $z=10$~mm. Layout same as in Figure~\ref{fig:5}.}
            \label{fig:8}
        \end{minipage}%
    }
\end{figure}

Figures~\ref{fig:5}--\ref{fig:8} present the results for multi-step rollout prediction across longitudinal projected, longitudinal slice and transverse slice modalities. Here, we compare three model configurations: (i) the baseline single-step training model, (ii) the multi-step rollout training model (5-step rollout), and (iii) the multi-step rollout training model with pushforward trick (5-step rollout with pushforward). All three models predict up to five consecutive time steps for comparison. For each modality, the error maps include mean absolute error (MAE) values for quantitative comparison. Notably, the MAE tends to increase for predictions generated by the multi-step rollout models (both with and without pushforward) compared to the baseline single-step training model, reflecting the accumulation of error over longer horizons. Nevertheless, qualitative inspection of the predicted fields reveals that the multi-step rollout models capture intrinsic flow details and fine-scale structures better than the baseline single-step training model, particularly in regions of complex mixing and internal gradients. The pushforward-trained model also qualitatively yields more coherent multi-step rollouts than configuration (ii). Overall, these results highlight the trade-off between short-term accuracy and long-term physical fidelity in autoregressive spatiotemporal modeling, and demonstrate the benefit of multi-step training for learning rich flow representations.

\subsection{Feature Transformation}
The feature transformation task assesses the model's ability to perform intra-timestep inference—learning deterministic mappings between different subsets of physical variables or between modalities. Unlike the spatiotemporal task, there is no temporal component here; the model processes data within a fixed time slice and outputs the corresponding transformed field. Separate models are trained for feature transformation tasks, sharing the same SwinV2-UNet Transformer backbone architecture but without time-step embeddings. The data source token is adjusted to encode the input modalities for each case. We explore five representative transformation test cases to assess model performance for cross-modal and cross-variable inference:

\noindent\textbf{Case 1:} Density $\rightarrow$ Velocity components ($u$, $v$, $w$). 
\textit{Objective:} Evaluate the model's ability to infer directional flow components from scalar density distributions, tested separately for longitudinal projected and longitudinal slice modalities.

\noindent\textbf{Case 2:} Longitudinal projected density $\rightarrow$ Transverse slice reconstruction. 
\textit{Objective:} Assess cross-dimensional reconstruction performance between longitudinal line-of-sight projections and transverse slice representations.

\noindent\textbf{Case 3:} Longitudinal projected density + velocity components $\rightarrow$ Longitudinal slice density + velocity components. 
\textit{Objective:} Evaluate the model's ability to fuse information from multiple projected fields (density and velocity components) to reconstruct local slice fields, effectively learning an inverse projection operation.

\noindent\textbf{Case 4:} Longitudinal slice density + velocity components $\rightarrow$ longitudinal projected density + velocity components. 
\textit{Objective:} Test the model's ability to aggregate localized slice information into a global projected representation.

\noindent\textbf{Case 5:} Transverse slice density at $z=2$~mm $\rightarrow$ Transverse slice density at $z=10$~mm. 
\textit{Objective:} Examine spatial transfer capability across distinct measurement planes.


These five test cases collectively span variable-to-variable, modality-to-modality, and spatial cross-plane transformations, offering a broad set of tests for the transformer architecture. Each case uses its own fitted weights on the curated simulation ensemble (Table~\ref{tab:curated_datasets}). The discussion and figures below show model outputs at $t=1.00$~s for the F--LES--IG realization as a representative instant; all Case~1--5 figures use this time step.

\subsubsection{Case 1: Density to Velocity Components}
\begin{figure}[htbp]
    \centering
    \includegraphics[width=1\textwidth]{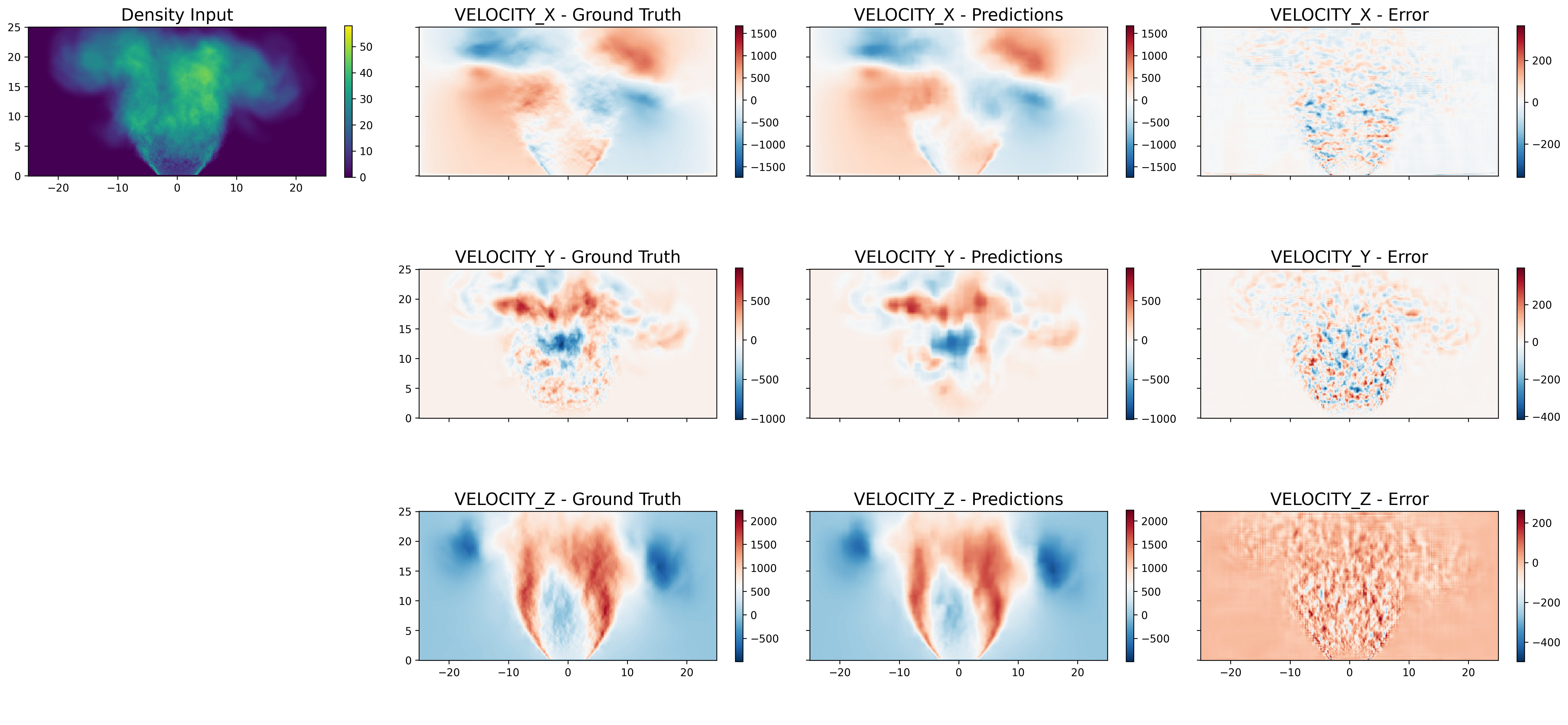}
    \caption{Feature transformation results for Case 1: longitudinal projected density to velocity components ($u$, $v$, $w$) using Euclidean loss.}
    \label{fig:9}
\end{figure}

\begin{figure}[htbp]
    \centering
    \includegraphics[width=1\textwidth]{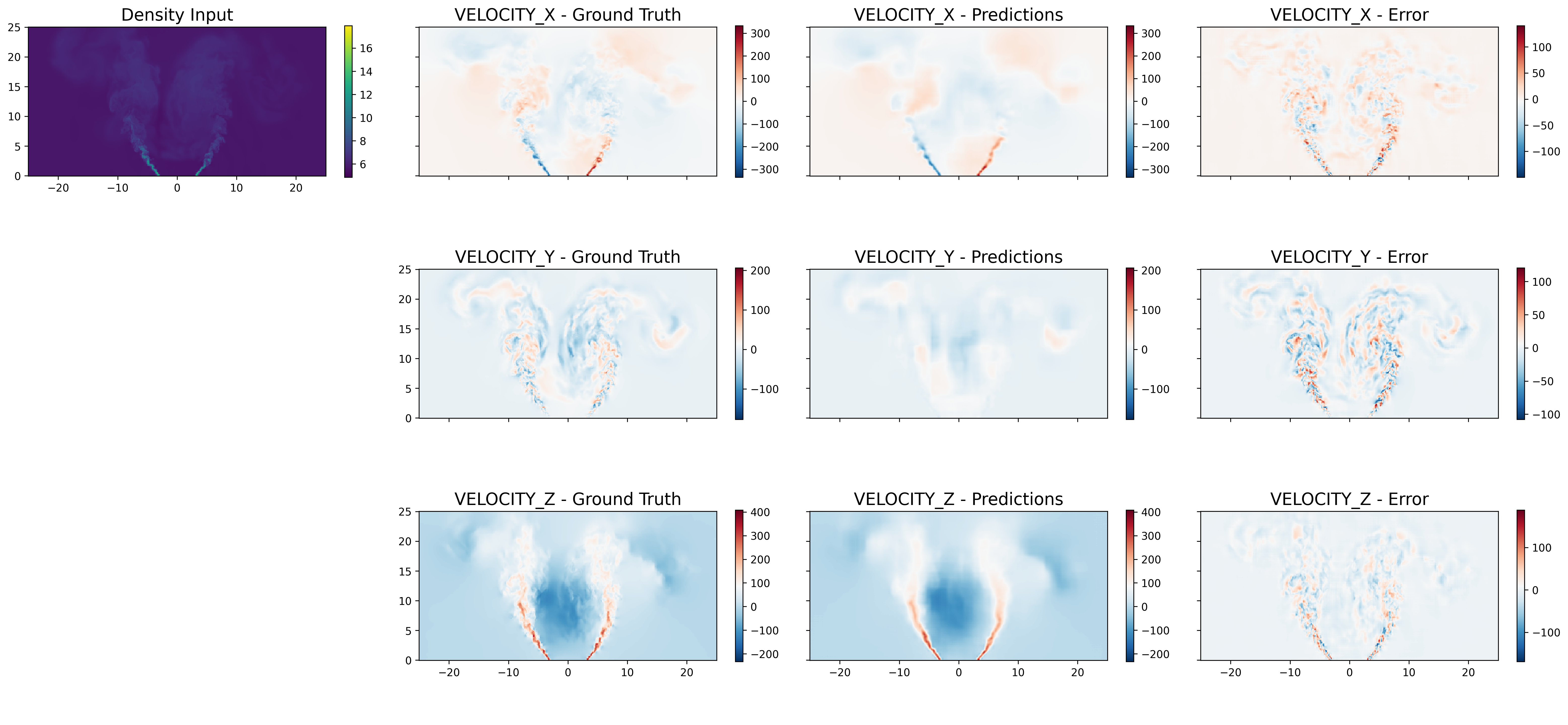}
    \caption{Feature transformation results for Case 1: longitudinal slice density to velocity components ($u$, $v$, $w$) using Euclidean loss.}
    \label{fig:10}
\end{figure}

Case 1 evaluates the model's capacity to infer velocity fields from density distributions. Figure~\ref{fig:9} presents results for longitudinal projected density to velocity component transformation, demonstrating the model's ability to extract directional flow information from line-of-sight integrated density fields. The predicted velocity components show that the $x$ \hide{component (streamwise direction) is well captured, and similarly the}and $z$ velocity components are accurately captured. However, the $y$ component, which is perpendicular to the observation plane, is not captured as well. This behavior is expected: the $y$ component represents out-of-plane motion that is inherently ambiguous when inferring from planar density distributions, as density gradients primarily encode in-plane flow structures. Figure~\ref{fig:10} shows analogous results for longitudinal slice-based density to velocity inference, where the model successfully reconstructs all three velocity components from planar density slices. The longitudinal slice-based transformation exhibits the same trend, with the $y$ component showing reduced accuracy compared to the $x$ and $z$ components. Overall, both qualitatively and quantitatively, the predictions appear better for the projected case than for slice. However, in both cases, intrinsic details are missing, as can be observed from the error plots, indicating a smoothing effect, which is also observed across all transformation cases discussed hereafter.

\subsubsection{Case 2: Longitudinal Projected Density to Transverse Slice Reconstruction}
\begin{figure}[htbp]
    \centering
    \includegraphics[width=1\textwidth]{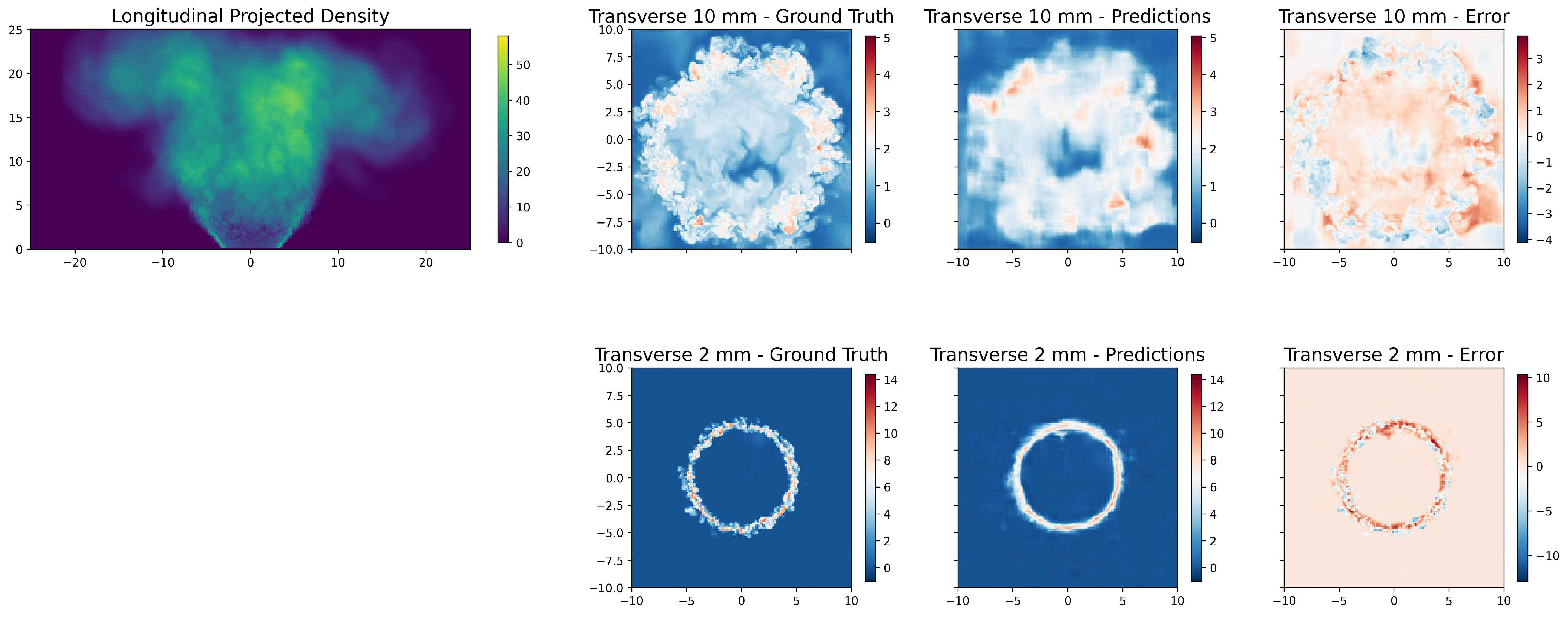}
    \caption{Feature transformation results for Case 2: longitudinal projected density to transverse slice density reconstruction.}
    \label{fig:11}
\end{figure}

Case 2 examines the model's ability to perform cross-dimensional reconstruction, transforming longitudinal line-of-sight projected density fields into corresponding transverse slice representations. Figure~\ref{fig:11} illustrates the reconstruction performance, showing that the model successfully infers transverse planar density distributions from cross-dimensional integrated projections. The transformation preserves key flow features including jet boundaries, mixing regions, and density gradients, demonstrating effective learning of the cross-dimensional projection-to-slice mapping. There is a lot of smoothing in this case though, more than what is seen in Case 1, which is expected given the inherent information loss in the projection-to-slice inverse problem. However, the model correctly picks up the timing when the transverse plane starts to show at $z=2$~mm, and although this cannot be shown in the static figure, temporal evaluation confirms that this timing is also correctly captured for both the $z=2$~mm and $z=10$~mm cases. The overall spatial structure and density distribution are accurately recovered, indicating robust cross-modal feature extraction capabilities despite the smoothing effects.

\subsubsection{Case 3 \& 4: Transformation between longitudinal - Slice and Projected Fields}
\begin{figure}[htbp]
    \centering
    \includegraphics[width=1\textwidth]{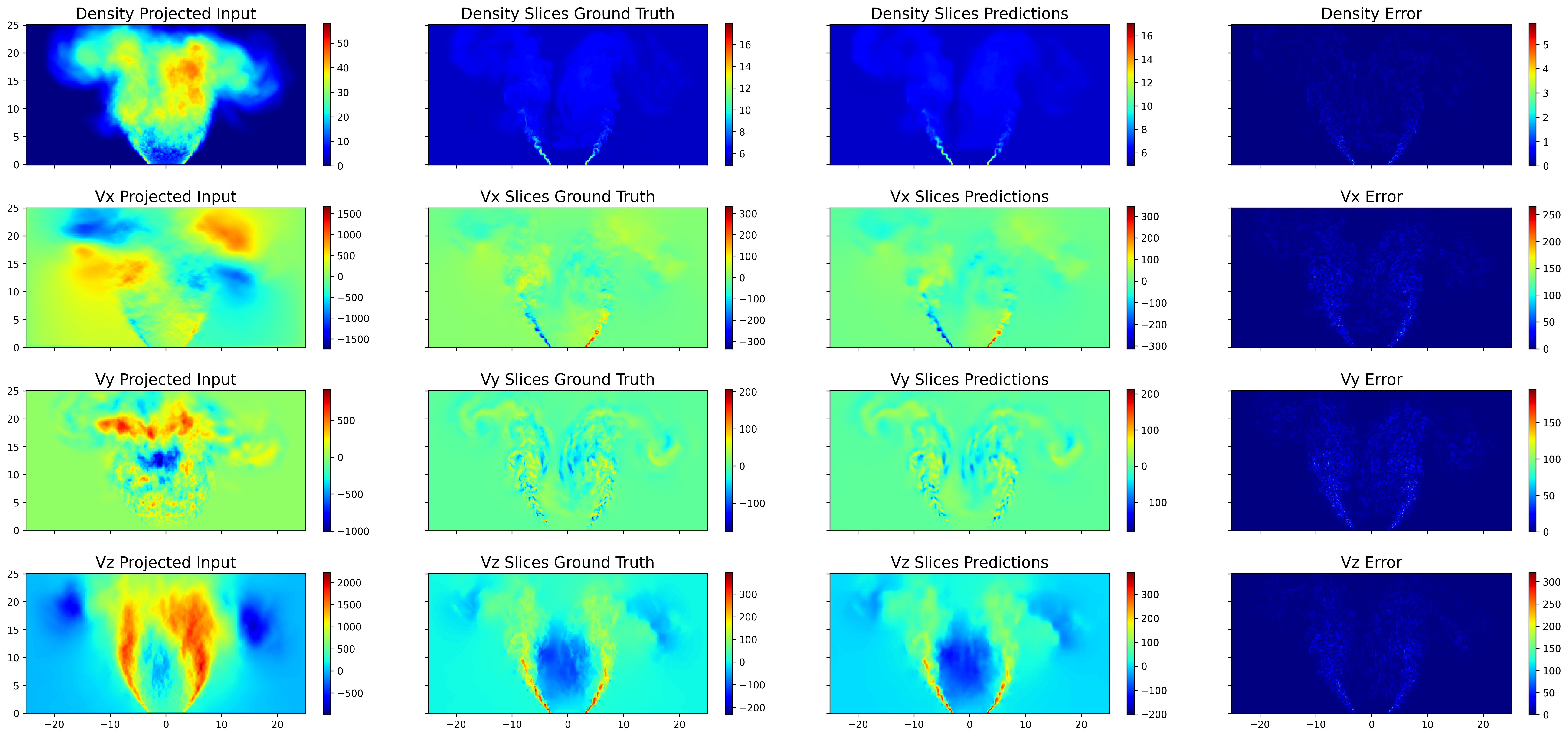}
    \caption{Feature transformation results for Case 3: longitudinal projected density + velocity components to longitudinal slice density + velocity components.}
    \label{fig:12}
\end{figure}

Cases 3 and 4 evaluate bidirectional transformation between slice and projected representations in longitudinal direction. Case 3 transforms projected density and velocity fields into corresponding slice representations, while Case 4 performs the inverse transformation, aggregating slice-based density and velocity information into projected representations. Figure~\ref{fig:12} demonstrates the model's performance for Case 3, where both density and velocity components are simultaneously inferred from projected inputs. The model successfully leverages the combined information from density and velocity projections to reconstruct accurate slice fields, with velocity components showing particularly strong agreement in regions of coherent flow structures. This capability to fuse information from multiple projected fields (multimodal fusion) enables reconstruction of local flow features from global line-of-sight measurements, which is valuable for diagnostic applications where only projected data are available.

\begin{figure}[htbp]
    \centering
    \includegraphics[width=1\textwidth]{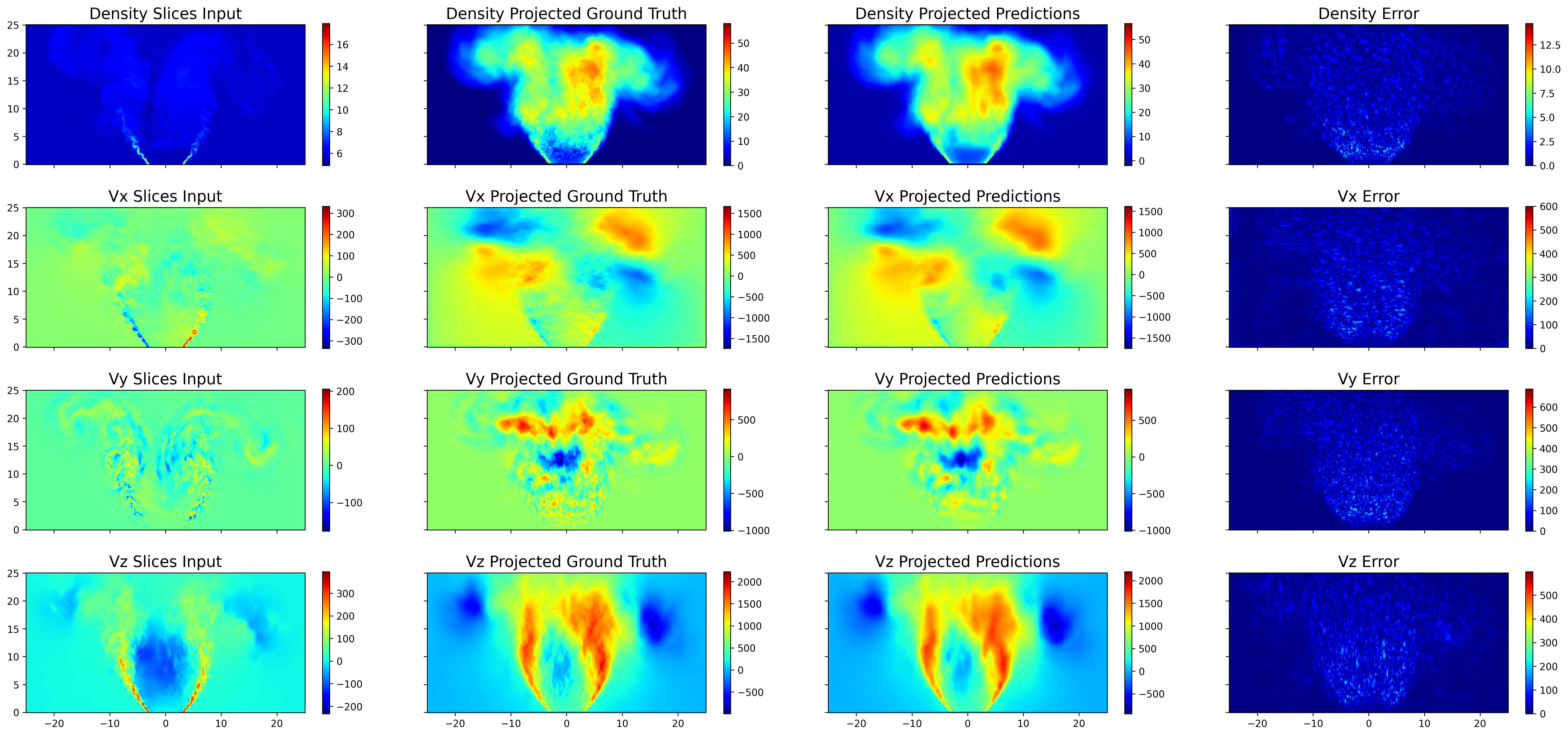}
    \caption{Feature transformation results for Case 4: longitudinal slice density + velocity components to longitudinal projected density + velocity components in longitudinal direction.}
    \label{fig:13}
\end{figure}

Figure~\ref{fig:13} shows the model's performance for Case 4, demonstrating its ability to transform localized slice representations into global projected fields, effectively performing a learned form of line-of-sight integration. The transformation preserves the large-scale flow structure while correctly aggregating slice information along the projection direction. This capability is particularly useful for synthesizing projected views from limited slice measurements, enabling efficient data fusion across different experimental modalities and measurement configurations.

\subsubsection{Case 5: Cross-Plane Spatial Transfer}
\begin{figure}[htbp]
    \centering
    \includegraphics[width=1\textwidth]{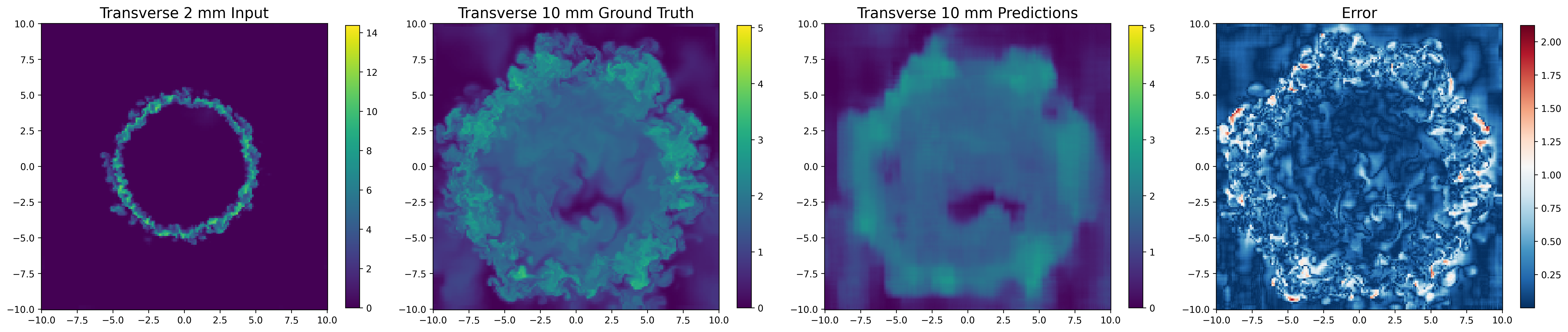}
    \caption{Feature transformation results for Case 5: transverse slice density at $z=2$~mm to transverse slice density at $z=10$~mm.}
    \label{fig:14}
\end{figure}

Case 5 examines spatial transfer across different axial planes, transforming transverse slice density at $z=2$~mm to corresponding slice at $z=10$~mm. Figure~\ref{fig:14} presents the cross-plane transformation results, demonstrating the model's ability to infer flow structures at downstream locations from upstream measurements. The transformation captures the expected flow evolution, including jet expansion and mixing enhancement between the two planes. There is a lot of smoothing in this case, more than what is seen in Case 1, which reflects the challenge of inferring downstream flow structures from upstream measurements. However, the model correctly picks up the timing when the transverse plane starts to show at $z=10$~mm, and although this cannot be shown in the static figure, temporal evaluation confirms that this timing is also correctly captured. While some fine-scale turbulent features differ due to the inherent flow variability, the overall flow topology and large-scale structures are accurately transferred, indicating successful learning of spatial flow correlations across different measurement planes.

\section{Conclusion}
In this work, a multimodal transformer-based modeling framework is presented for prediction of complex fluid flows relevant to energy systems, built on a SwinV2 Vision Transformer-based encoder-decoder architecture. The framework addresses two complementary tasks: autoregressive spatiotemporal prediction for forecasting flow evolution, and feature transformation for inferring unobserved fields or modalities from limited measurements. By training on a diverse dataset of multi-fidelity CFD simulations spanning multiple grid resolutions, turbulence models, and thermodynamic treatments, the models learn to generalize across different physical regimes and data modalities while maintaining a unified architecture.

The results demonstrate that the transformer-based model successfully captures large-scale flow structures and temporal evolution across longitudinal and transverse representations. For spatiotemporal prediction, the model accurately forecasts flow-fields over multiple time steps, with multi-step rollout training configurations capturing intrinsic flow details better compared to single-step training, despite increased error accumulation over longer horizons. The model's ability to preserve sharp interfaces and global motion is robust, though recovery of fine-scale turbulent features remains challenging, consistent with the smoothing effects observed across all transformation tasks.

The feature transformation results reveal both capabilities and limitations of the framework. The model successfully infers velocity components from density distributions, with strong performance for in-plane velocity components ($x$ and $z$) but reduced accuracy for out-of-plane motion ($y$ component), reflecting the inherent ambiguity in inferring three-dimensional flow structures from planar observations. Cross-modal transformations between longitudinal projected and slice representations demonstrate effective learning of projection-to-slice mappings, though with significant smoothing effects. The model correctly captures temporal aspects of flow development, including the timing of transverse plane appearances, indicating successful learning of spatiotemporal correlations.

This work establishes a proof-of-concept for applying vision transformer-based models to fluid dynamics problems relevant to energy technologies, demonstrating how training on diverse simulation datasets can enable models that generalize across resolutions, turbulence models, and observational modalities. Future work will extend this framework by scaling the current model architecture with highly efficient window-sequence-pipeline parallelism techniques such as SWiPe~\cite{hatanpaa2025aeris}. We will also explore incorporation of advanced flow matching and latent masked training approaches such as OmniCast~\cite{nguyen2025omnicast} to enable efficient probabilistic modeling. Lastly, model applicability to complex geometries and mesh topologies, typical for energy system configurations and datasets, will be improved by replacing patch-based data representation with graph-based and point cloud-based representations. These efforts will contribute to building large-scale scientific foundation models for energy systems-relevant fluid dynamics applications. The framework provides a foundation for developing fast, data-driven surrogate models that can complement or replace computationally expensive simulations in engineering design and optimization workflows.

\section*{Acknowledgments}
The manuscript has been created by UChicago Argonne, LLC, Operator of Argonne National Laboratory (Argonne). The U.S. Government retains for itself, and others acting on its behalf, a paid-up nonexclusive, irrevocable world-wide license in said article to reproduce, prepare derivative works, distribute copies to the public, and perform publicly and display publicly, by or on behalf of the Government. This work was supported by the U.S. Department of Energy (DOE), Office of Science under contract DE-AC02-06CH11357. The authors acknowledge Laboratory-Directed Research and Development (LDRD) funding support from Argonne’s Advanced Energy Technologies (AET) directorate. Lastly, the authors are thankful to Drs. Riccardo Scarcelli and Diego Bestel from Argonne National Laboratory for sharing the simulation datasets used in this work.

\bibliographystyle{unsrt}
\bibliography{references}

\clearpage

\clearpage

\clearpage

\clearpage

\end{document}